\documentclass[11pt,tightenlines,eqsecnum,floats,aps,amsmath,amssymb,nofootinbib,prd,shownopacs,floatfix]{revtex4}

\usepackage{epstopdf}
\usepackage{latexsym}
\usepackage{graphicx}
\usepackage{amsmath}
\usepackage{amssymb}
\usepackage{mathrsfs}
\usepackage{xparse}
\usepackage{mathtools}
\usepackage{wasysym}
\usepackage{fancyhdr}
\usepackage[utf8]{inputenc}
\usepackage{mathrsfs}
\usepackage{soul}
\usepackage{float}
\usepackage[center]{subfigure}
\usepackage{color}
\usepackage{xparse}
\usepackage{appendix}

\begin{document}
\title{ Tunneling wavefunction proposal with loop quantum geometry effects }

\author{Meysam Motaharfar}
\email{mmotah4@lsu.edu}

\author{Parampreet Singh}
\email{psingh@lsu.edu}
\affiliation{Department of Physics and Astronomy, Louisiana State University, Baton Rouge, LA 70803, USA}


\begin{abstract}
In Vilenkin's tunneling wavefunction proposal our expanding universe is born via a tunneling through a barrier from nothing at the zero scale factor. We explore the viability of this 
proposal for the spatially closed FLRW model  with a positive cosmological constant including quantum gravity modifications in the Planck regime. Our setting is the effective spacetime description of loop quantum cosmology (LQC) which is known to replace the big bang singularity with a bounce due to the holonomy modifications. Due to the bounce, the barrier potential of the Wheeler-DeWitt theory is replaced by a step like potential which makes the tunneling proposal incompatible. But for a complete picture of singularity resolution, inverse scale factor modifications from quantum geometry must be included which play an important role at very small scale factors in  the spatially closed models. We show that with inclusion of inverse scale factor modifications the resulting potential is again a barrier potential.  The universe at the vanishing scale factor is dynamically non-singular and in an Einstein static like phase. We show that quantum geometric effects in LQC provide a non-singular completion of Vilenkin's tunneling proposal. We also find that quantum geometric effects result in a possibility of a tunneling to a quantum cyclic universe albeit for a very large value of cosmological constant determined by the quantum geometry.

\end{abstract}

\maketitle

\section{Introduction}
One of the oldest questions in cosmology is whether the universe had a beginning or was past-eternal. The singularity theorems of Penrose, Hawking and Geroch provide a rigorous answer to this question in general relativity (GR) proving that under certain reasonable assumptions, 
with energy conditions as a key ingredient, the universe must begin from a singularity in the past \cite{Geroch:1968ut,Hawking:1970zqf,Hawking-Ellis}. An implication of these theorems is that in the cosmological dynamics if the weak energy condition is violated then one can in principle resolve the big bang singularity. Though the process generally requires a very careful choice of initial conditions and/or a fine tuning of parameters, bypassing singularity theorems via violation of energy conditions opened a window  to construct non-singular models. As an example, it was believed that inflationary models can be past eternal if weak energy condition can be violated due to quantum effects. 
However, Borde, Guth, and Vilenkin (BGV) using just kinematic arguments and assuming that the universe was on average expanding in the past, proved that inflationary cosmology is past-incomplete \cite{Borde:2001nh}. Although there exist counter-examples of the BGV theorem, such as the emergent universe scenario \cite{Ellis:2002we} and the oscillating universe \cite{ Graham:2011nb}, where the average expansion is zero, it has been shown that such models even if they can be built without fine-tuning, are quantum mechanically unstable \cite{ Mithani:2011en, Mithani:2012ii, Mithani:2014jva}. It turns out that in classical gravity, it is difficult to construct cosmological models which can bypass BGV theorem and be internally consistent. Thus, it is inescapable that a classical universe had a beginning in the past. But classical GR is an incomplete theory where the singularities act as the boundaries of classical spacetime where the geodesic evolution ends. It has been long expected that a quantum theory of gravity would result in important insights on the question of resolution of cosmological singularities, past incompleteness as well as the boundary conditions and the dynamical laws valid at the birth of our universe.

These fundamental questions about the boundary conditions and the wavefunction of our universe  have been explored in detail and debated in quantum cosmology. 
Two popular proposals to specify the boundary condition of the wavefunction of the universe 
are  the Vilenkin's ``{\it tunneling proposal}" \cite{Vilenkin:1982de, Vilenkin:1984wp} and Hartle-Hawking's ``{\it no-boundary proposal}" \cite{Hartle:1983ai, Hawking:1983hj}. In a one-dimensional minisuperspace setting 
the underlying physics of both of these proposals can be understood  via a quantum tunneling through a classical barrier \cite{Vilenkin:1987kf}. An example of such a setting arises in a  spatially closed FLRW universe with a positive cosmological constant. In this model,  a barrier resulting from an interplay of the intrinsic curvature and the positive cosmological constant has a form determined from an   effective minisuperspace potential in the Wheeler-DeWitt equation. Recall that the Wheeler-DeWitt equation is the quantum Hamiltonian constraint which in the classical theory yields the Friedmann equation. In the Wheeler-DeWitt theory, states which are sharply peaked in a macroscopic universe at late times when evolved backwards towards the big bang follow the classical trajectories determined from Friedmann dynamics to a great accuracy when the quantum fluctuations remain small throughout the evolution \cite{Ashtekar:2006uz}. Due to this correspondence,  as in the classical mechanical problems, the allowed and forbidden regions of the  barrier obtained from the   effective minisuperspace potential in the Wheeler-DeWitt equation can also be determined using the classical dynamical equations, in the present case the Friedmann equation for the $k=1$ FLRW model. More precisely, the   effective minisuperspace potential determined from the Wheeler-DeWitt equation can also be extracted from the Friedmann equation using an overall scaling by a term proportional to a power of scale factor.

In the classical theory, for a spatially closed FLRW model sourced with a positive cosmological constant, the universe contracting from a very large size bounces at turnaround radius $a = \sqrt{3/\Lambda}$, and then re-expands. The region below this scale factor is classically forbidden if no other energy or matter content is present.  
If we consider the analogy of tunneling through the barrier for both boundary proposals, in Vilenkin's tunneling proposal the wavefunction is composed of a sub-dominant growing and a dominant decaying mode under the barrier, and a spatially compact universe spontaneously nucleates out of a barrier. The tunneling wave function at large scale factor can be seen as just ``outgoing" waves (expanding universe), like a wave function of a particle escaping a radioactive nucleus. An expanding universe is born from `nothing' where ‘nothing’ refers to a state with no classical space and time at the big bang \cite{Vilenkin:1987kf}. On the other hand, in the Hartle-Hawking's no-boundary proposal, 
the wavefunction is a fine-tuned superposition of ingoing and outgoing waves (contracting and expanding universe) with equal amplitudes such that wave function is real and decreases towards big bang singularity under the barrier. For a given positive cosmological constant $\Lambda$, the amplitude of wavefunction  scales as  
\begin{equation}\label{VHH}
\Psi \sim \exp \left(\pm\frac{c}{\Lambda}\right)    
 \end{equation}
with $c$ being a positive constant and a positive sign in the exponential for the no-boundary proposal, and a negative sign for the Vilenkin's proposal of tunneling. 
Since the amplitude for tunneling wavefunction peaks at a large value of cosmological constant, it prefers tunneling to smaller expanding universe. In contrast, the amplitude of the wavefunction is larger for a smaller cosmological constant in case of the no-boundary proposal which means it prefers tunneling to a larger expanding universe.

Above analogy between the tunneling wavefunction proposal and the no-boundary wavefunction breaks down in the  higher dimensional minisuperspace  setting and one resorts to the 
path integral formulation of these boundary proposals \cite{Fanaras:2022twv}. In the path integral formulation, it is conjectured that tunneling wavefunction can be expressed as a path integral over Lorentzian histories interpolating between a vanishing 3-geometry and a given configuration in the superspace \cite{Vilenkin:1984wp}
\begin{align}
    \Psi_{V} = \int_{\emptyset}^{(g, \phi)} \mathcal{D}g\mathcal{D}\phi \, e^{iS},
\end{align}
while the Hartle-Hawking wavefunction is defined as a path integral over compact Euclidean histories bounded by a given 3-geometry and matter field configuration \cite{Hartle:1983ai}

\begin{align}
    \Psi_{HH} = \int^{(g, \phi)} \mathcal{D}g\mathcal{D}\phi \, e^{-S_{E}},
\end{align}
where $S_{E}$ is the Euclidean action obtained by the standard Wick rotation $t \rightarrow -i \tau$. There have been interesting developments in this setting recently. 
Using Lorentzian path integral and Picard-Lefschetz theory, a saddle point analysis shows  that the no-boundary proposal results in  the same prediction as the tunneling wavefunction \cite{Feldbrugge:2017kzv} whose implications including perturbations have been studied in detail \cite{Feldbrugge:2017fcc, DiazDorronsoro:2017hti, Feldbrugge:2017mbc, DiazDorronsoro:2018wro, Feldbrugge:2018gin, Vilenkin:2018dch,Vilenkin:2018oja, deAlwis:2018sec, DiTucci:2019dji}. Despite these remarkable results, above mentioned works are based on using a semi-classical description of gravity while resolving the singularity by closing off the geometry at the bottom through a Euclidean continuation. 
Hence, it is pertinent to ask in what way these predictions about the initial state of the universe, and as a result the tunneling and the no-boundary proposals might be affected when one includes non-perturbative quantum gravity effects resulting in a non-singular dynamics. One way to answer this question is via the modifications to the effective minisuperspace potential which can be obtained if the details of the modified dynamics near the classical singularity are available.

An arena to understand effects of underlying quantum gravity on these proposals which allows to understand quantum geometry effects via modifications to the effective minisuperspace potential  is loop quantum cosmology 
(LQC) \cite{Ashtekar:2011ni} which is a canonical quantization of cosmological spacetimes using techniques of loop quantum gravity (LQG) for homogeneous spacetimes. Here 
 quantum geometry results in a generic resolution of singularities in isotropic as well as anisotropic spacetimes \cite{Singh:2009mz,Singh:2010qa,Singh:2011gp,Singh:2014fsy,Saini:2017ipg,Saini:2017ggt,Saini:2016vgo}. Quantum gravitational effects in LQC for spatially curved models can arise in  two ways: via holonomies of the connection variables which lead to modifications when the spacetime curvature becomes Planckian, and 
also through modifications to the  inverse scale factor terms. The latter become prominent only close to Planck length.
In the cosmological context, the Wheeler-DeWitt equation is replaced by a quantum difference equation arising from the loop quantization of the Hamiltonian constraint. The underlying discrete quantum geometry is directly responsible for an upper bound on the spacetime curvature causing a non-singular  bounce when energy density of the matter content reaches a maximum value 
\cite{Ashtekar:2006rx,Ashtekar:2006wn}.  For spatially-flat model the expectation value of volume is bounded above zero in the physical Hilbert space \cite{Ashtekar:2007em} and the probability of bounce at non-zero volume is unity \cite{Craig:2013mga}. In the presence of spatial curvature, holonomy modifications play the most significant role in singularity resolution \cite{Ashtekar:2006es}, nevertheless the role of inverse scale factor modifications can become important especially when spatial curvature and anisotropies are present together \cite{Gupt:2011jh}. 
The non-singular bounce has been rigorously confirmed in various models using high performance computing \cite{Diener:2014mia, Singh:2018rwa}, including anisotropic vacuum spacetimes \cite{Diener:2017lde}, and these studies reveal that one can capture the underlying quantum discrete evolution in an effective spacetime description where quantum geometry effects are encoded in an effective Hamiltonian constraint \cite{Taveras:2008ke} which allows to obtain a modified Friedmann equation with quantum gravity corrections \cite{Singh:2006sg}.\footnote{The usage of ``effective'' in LQC part of this manuscript should not be confused with the ``effective potential'' obtained either from the  Wheeler-DeWitt equation or the Friedmann dynamics in classical theory or LQC. For both, the Wheeler-DeWitt theory and LQC, the potential is labeled as ``effective potential.'' } Interestingly, the modified Friedmann equation captures the the underlying quantum evolution to an excellent accuracy if the quantum fluctuations are small \cite{Diener:2014mia, Singh:2018rwa}. Using this modified Friedmann dynamics one can obtain the quantum geometric modifications to the effective minisuperspace potential, and analyze the affects on the tunneling proposal.

In the bouncing models in quantum cosmology, one often interprets the singularity resolution as arising from some sort of repulsive character of modifications to classical gravity. As an example, in some works the repulsive force near the classical singularities can be obtained from effective minisuperspace potential which diverges to infinity near the zero scale factor \cite{Hertog:2021jyd, Matsui:2021yte, Martens:2022dtd}. The big bang is protected by an infinite hard-wall wall in such a case. It turns out that there is a similar, but a finite wall or a step-like potential in LQC if one considers holonomy modifications. The situation is similar to a hard wall problem in ordinary quantum mechanics as a result of which the wave function of universe should vanish at the big bang and the decaying mode is no longer dominant. This seems to put the tunneling wavefunction proposal at an incompatible footing with such bouncing models. With the above analogy one may conclude that  the no-boundary proposal is favored to explain the  beginning of the universe.\footnote{Having infinite barrier is also consistent with DeWitt boundary condition which requires wavefunction of the universe vanishes at zero scale factor, i.e. $\psi\left(a=0\right)=0$. } But this picture, at least in LQC, as we will see is far from complete unless one takes into account all potential quantum gravitational modifications which can potentially change the details of singularity resolution.

In this manuscript our goal is to understand the above issue taking into account the role of inverse scale factor modifications which must be included to understand the complete picture of singularity resolution in the presence of spatial curvature in LQC. To understand  the quantum geometric modifications to the   effective minisuperspace potential we use the modified Friedmann equation in LQC. We assume its validity throughout the evolution. When only holonomy modifications are considered the bounce happens at a non-vanishing scale factor and the resulting   effective minisuperspace potential turns out to be a wall with a large Planckian magnitude. This makes tunneling proposal incompatible as in other bouncing models. However, when we consider a more complete picture of singularity resolution by including  inverse scale factor modifications then the effective minisuperspace potential near $a=0$ is modified in such a way that a quantum barrier as in Wheeler-DeWitt appears but with a much larger magnitude. In contrast to the Wheeler-DeWitt theory the underlying quantum evolution and modified Friedmann dynamics is non-singular. The universe in LQC can tunnel from $a=0$ where both $\dot a$ and $\ddot a$ vanish and there is no singularity. As a result, we find that quantum geometric effects in LQC  through inverse scale factor modifications actually make the  tunneling wavefunction proposal complete by resolving its singularity at $a=0$. It also turns out that there exists a critical value of cosmological constant above which the universe can tunnel to a quantum cyclic universe. This value determined by quantum geometry is quite large, $\Lambda > 10.2887$ in Planck units. 
It is important to note a caveat of our analysis. The inverse scale factor effects only become important in the deep Planck regime when the scale factor approaches values close to Planck length. In this regime one expects fluctuations to be large and one might suspect the validity of effective spacetime description. However, it turns out that the effect of large fluctuations translates to a lower bounce density \cite{Corichi:2011rt,Diener:2014hba} and even in such cases the form of the modified Friedmann equation does not change except that the bounce density decreases \cite{Ashtekar:2015iza}. One expects these results to hold true also for $k=1$ model, and if so this would amount to a decrease in the height of the barrier potential without qualitatively affecting any results. 


The outline of the paper is as follows. In Sec. \ref{Section II}, we will review boundary proposals and discuss how to obtain  effective minisuperspace potential from Friedmann equation in classical cosmology. In Sec. \ref{Section III}, we will obtain the effective Friedmann and Raychaudhuri equations containing both holonomy and inverse scale factor corrections and discuss the fate of the universe at $a=0$ which turns out to be non-singular. Then, we  find the effective minisuperspace potential from effective Friedmann equation and discuss implications of quantum geometry effects for tunneling wavefunction proposal in Sec. \ref{Section IV}. Here we note differences in potential when one just includes the holonomy modification and when one includes both. We discuss these cases for different values of cosmological constant and show that it is possible to tunnel to a large expanding universe as well as to a cyclic quantum universe. We conclude with a summary  in Sec. V. We use Planck units in all the figures.

\section{Wheeler-DeWitt Quantum Cosmology and boundary proposals}\label{Section II}

In this section we briefly review Wheeler-DeWitt quantum cosmology, boundary proposals, the no-boundary and the tunneling wavefunctions, and finally discuss how to obtain  the effective minisuperspace potential from Friedmann equation in classical cosmology. We consider a spatially closed universe sourced with a positive cosmological constant given by the following action
\begin{align}\label{action}
S = \frac{1}{2\kappa} \int \sqrt{-g} \, \mathrm{d}^{4}x  \left( R - 2{\Lambda}\right),
\end{align}
where $\kappa = 8 \pi G$ and $\Lambda$ is cosmological constant. Considering that the universe is isotropic and homogeneous, the metric is given by
\begin{align}\label{metric}
\mathrm{d} s^2 = - N^2(t) \mathrm{d} t^2 + a^2(t) \mathrm{d} \Omega_{3}^2,
\end{align}
in which $a(t)$ is scale factor, $N(t)$ is lapse function and $d\Omega_{3}$ is the metric on unit 3-sphere. Inserting metric (\ref{metric}) into action (\ref{action}) and considering $\dot N =0$, one can find the following Lagrangian density  
\begin{align}\label{lagrangian}
\mathcal{L} = \frac{1}{\kappa} \left(\frac{-3 a\dot a^2}{N} +3 {N a}- N a^3 {\Lambda}\right) . 
\end{align}
While taking derivative with respect to lapse function from Lagrangian (\ref{lagrangian}) and choosing $N=1$ results in Friedmann equation 
\begin{align}\label{Friedmann}
 H^2 = \frac{\Lambda}{3} - \frac{1}{a^2},
\end{align}
with $H =\dot a/a$ being Hubble parameter. The classical solution for  Eq. (\ref{Friedmann}) is de Sitter space, i.e., $a = \sqrt{{3}/{\Lambda}} \cosh(\sqrt{\Lambda/3}~  t)$ which means that the universe starting from an infinite size, contracts, bounces and re-expands. See Fig. 1 for the plot of the effective minisuperspace potential which denotes the bounce point (B) at $a = \sqrt{3/\Lambda}$. Note that the classical Friedmann dynamics, as well as the Wheeler-DeWitt equation, is singular at $a=0$. The singularity at the vanishing scale factor is separated from the classical bounce point by a barrier. 

\begin{figure}
    \centering
    \includegraphics[scale = 0.6]{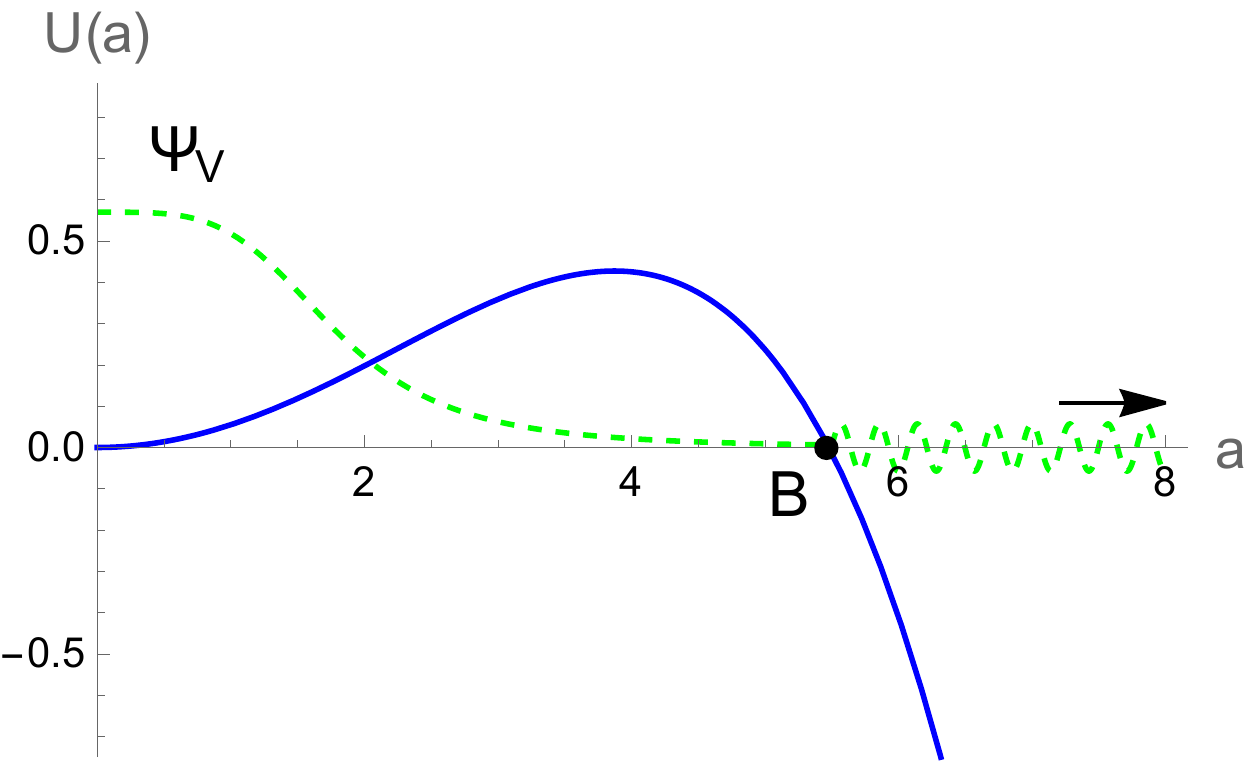}
     \includegraphics[scale = 0.6]{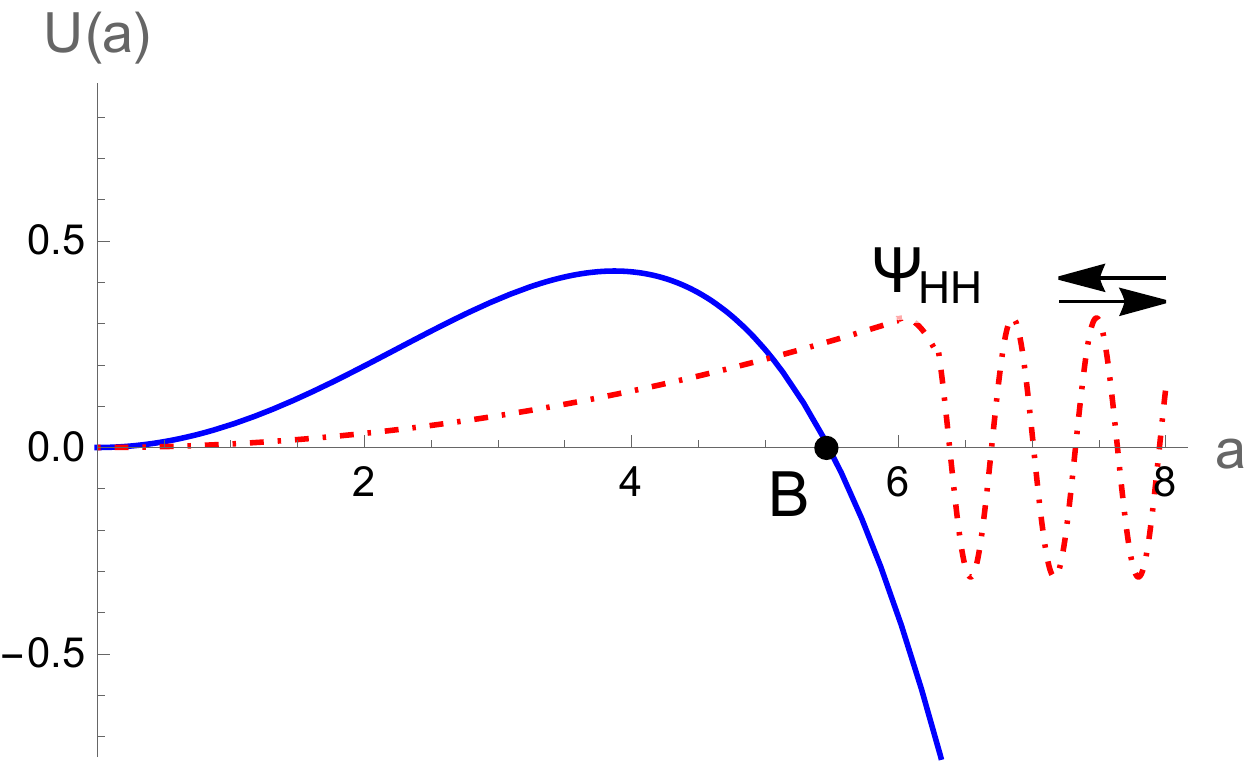}
    \caption{Schematic behavior of tunneling (left) and no-boundary (right) wavefunction proposals in Wheeler-DeWitt quantum cosmology. Point $B$ indicates the bouncing turnaround point in classical cosmology for a large contracting universe. We set $\Lambda=0.1$ in Planck units.}
    \label{Fig1}
\end{figure}

The Hamiltonian constraint for this minisuperspace model can be obtained using Lagrangian density (\ref{lagrangian}), which is given by
\begin{align}\label{eq:Ham-cons-cl}
    \mathcal{H} = - \frac{\kappa}{12a}\left[ {p_{a}^2} +  \frac{36a^2}{\kappa^2} \left(1 -{\frac{\Lambda}{3}}a^2\right)\right]=0,
\end{align}
where $p_{a} = - {6} a \dot a/ {\kappa}$ is conjugate momentum of scale factor. Upon quantization, by replacing $p_{a} \rightarrow - i {\mathrm{d}}/{\mathrm{d}a}$, one obtains the Wheeler-DeWitt equation  
\begin{align}\label{Wheeler-DeWitt}
    \left [a^{-n} \frac{\mathrm{d}}{\mathrm{d}a} a^{n} \frac{\mathrm{d}}{\mathrm{d}a} - U(a)\right] \Psi(a) = 0,
\end{align}
where the parameter $n$ represents factor ordering ambiguity and $U(a)$ is the effective minisuperspace potential
\begin{align}\label{V}
U(a) = \frac{36}{\kappa^2} a^2 \left(1-  \frac{\Lambda}{3}a^2\right) .
\end{align}
From Eq. (\ref{Wheeler-DeWitt}) one can immediately see that it is Schrodinger equation with potential $U(a)$ given by Eq. (\ref{V}) and zero energy eigenvalue. From Fig. \ref{Fig1} in which we plot the effective minisuperspace potential (solid blue curve), one sees that the  effective minisuperspace potential has two classical regimes for zero energy which corresponds to the Hamiltonian constraint \eqref{eq:Ham-cons-cl}; a single point at zero scale factor, i.e., $a=0$, and for scale factor larger than bouncing turnaround point, i.e., $a>  \sqrt{{3}/{\Lambda}}$. Hence, Wheeler-DeWitt quantum cosmology is analogous to quantum mechanical problem with potential barrier in which the universe can start from zero size and zero energy, which means no classical spacetime and matter, and tunnel through the barrier to a classically expanding universe. As we discussed in the introduction, although no-boundary and tunneling wavefunction proposals were formulated in different way, they can be understood  using this quantum tunneling analogy \cite{Vilenkin:1987kf}. In the left panel of Fig. \ref{Fig1}, the dashed (green) curve illustrates the Vilenkin's wavefunction in which the wavefunction decreases under the barrier and it has just outgoing wave mode in classical region similar to wave function of a particle escaping a radioactive nucleus. The sub-dominant decaying mode is not shown for visual clarity. While the dashed (red) curve in the right panel of Fig \ref{Fig1} illustrates the no-boundary proposal in which the universe is in superposition of ingoing (contracting universe) and outgoing (expanding universe) wave modes in classical region, i.e., $a>\sqrt{3/\Lambda}$ in such a way that the wavefunction decreases towards the big bang singularity under the barrier. So the corresponding nucleation probability for these two proposals is given by modulus of wavefunction in Eq. (\ref{VHH}). %

%

In summary, having the  effective minisuperspace potential one can discuss the boundary conditions for the wavefunction of the universe in a qualitative manner. Although we obtained the effective minisuperspace potential from the Hamiltonian constraint and its corresponding Wheeler-DeWitt equation, one can also see that  effective minisuperspace potential can be obtained from the classical Friedmann equation. In fact, multiplying both sides of Friedmann equation by ${-36} a^4/{\kappa^2}$, the right hand side is just the effective minisuperspace potential term $U(a)$. 
The importance of this correspondence is that it allows us to capture the effective minisuperspace potential with quantum geometric effects if a modified Friedmann equation containing quantum gravity effects is known. In the following section we first summarize the modified Friedmann dynamics for the spatially closed model in LQC and then consider the implications of the resulting effective minisuperspace potential.


\section{Effective dynamics in k=1 Loop quantum cosmology}\label{Section III}

The canonical quantization in LQG is based on using Ashtekar-Barbero variables: the connection $A^i_a$ and conjugate triads $E^a_i$. In LQC, 
one performs a symmetry reduction of these variables at the classical level and then expresses the Hamiltonian constraint in terms of the  holonomies of the symmetry reduced connection $c$, and the symmetry reduced triads $p$. For the $k=1$ FLRW spacetime, the manifold is $\Sigma \times \mathbb{R}$, where the spatial manifold $\Sigma$ has a topology of a three sphere $\mathbb{S}^3$. This unit sphere has a physical volume $V = |p|^{3/2} = 2 \pi^2 a^3$, where as before $a$ denotes the scale factor of the universe. The connection and the triad variables satisfy 
\begin{equation}
\{c,p\} = \frac{\kappa \gamma}{3}
\end{equation}
where $\gamma$ is the Barbero-Immirzi parameter whose value is fixed by the black hole thermodynamics in LQG. As in various other works in LQC, we will fix this value as $\gamma = 0.2375$. While the triad is related to the square of the scale factor through a kinematical relation, the connection is a measure of the time derivative of the scale factor but this relation needs to be determined using Hamilton's equation. In the classical theory, up to the contribution from the intrinsic curvature $c$ is proportional to the extrinsic curvature, but when the quantum geometric effects in the
Hamiltonian are included this relation becomes much more non-trivial.

A key difference between LQC and the Wheeler-DeWitt theory is that at the quantum level the Hamiltonian constraint in LQC is a not a differential operator in volume representation. Due to the underlying quantum geometry, it turns out to be a difference operator with a uniform step in the volume. Note that the lattice on which the difference operator has support includes the vanishing volume. The quantum geometric modifications enter the Hamiltonian constraint in two distinct ways. The first is by expressing the field strength of the connection in terms of the holonomies of the connection which are computed over a loop with a minimum area determined by the quantum geometry. The second is via expressing the inverse scale factor terms in the Hamiltonian in terms of a Poisson bracket between the holonomies and triads. The first modification results in a non-local curvature operator, while the second qualitatively modifies the behavior of inverse scale factor near the Planck scale. In the following, as is the convention in LQC literature, we will denote the first modification as holonomy based and the second as inverse scale factor  based modifications. 

If we consider a spatially closed model sourced with a massless scalar field, then using quantum Hamiltonian constraint of LQC  the backward evolution of states peaked in a large classical universe result in a quantum bounce \cite{Ashtekar:2006es}. Note that this quantum bounce is of complete different origin than the classical bounce discussed in the previous section. As in other approaches, there are quantization ambiguities in the Hamiltonian constraint but the singularity resolution is a robust phenomena  \cite{Corichi:2011pg, Dupuy:2016upu} including for different types of matter \cite{Dupuy:2019ibu,Gordon:2020gel,Motaharfar:2021gwi}. At late times the quantum dynamics approximates the Wheeler-DeWitt evolution and classical GR is recovered. At late times, when the closed universe recollapses while the Wheeler-DeWitt evolution results in a big crunch singularity, LQC results in another quantum bounce leading to a non-singular cyclic evolution. It turns out that for values of scalar field momentum which result in a large universe the inverse scale factor modifications remain subdominant. The bounce is therefore often attributed as resulting from the holonomy modifications. However, the inverse scale factor  modifications by themselves can also result in a singularity resolution in LQC \cite{Singh:2003au}, and  the effect of these terms is effectively to make energy density vanishing as the scale factor approaches zero \cite{Singh:2005km}. The important point to note is that generally in numerical simulations carried out in LQC the role of inverse scale factor  modifications is masked by the holonomy modifications if one starts from a macroscopic universe which generally bounces at a scale factor much larger than the Planck length. But if one is interested in understanding the quantum geometry near $a=0$ the role of inverse scale factor modifications is quite important as will become evident from the effective potential. 


Interestingly the quantum dynamics in LQC can be captured accurately using an effective Hamiltonian constraint which taking into account both holonomy and inverse scale factor corrections is given by  \cite{Ashtekar:2006es}
\begin{align}\label{constraint}
\mathcal{H}_{eff} = \frac{A(v)}{16\pi G} \left[\sin ^2 \bar\mu \left(c- k\right) - k \chi \right] + \mathcal{H}_{M},
\end{align}
where $\mathcal{H}_{M} = \rho V$ is the matter Hamiltonian and 
\begin{align}
 \chi & := \sin^2 \bar \mu - (1+\gamma^2) \bar \mu^2, \\
\bar \mu^2 p &= 4 \sqrt{3} \pi \gamma l_{Pl}^2 := \Delta,
\end{align}
where $\Delta$ is the minimum area gap obtained from LQG. In this manuscript, we will assume that the effective spacetime description is valid at all the scales. The holonomy modifications are contained in the trigonometric terms, while the inverse scale factor modifications result in $A(v)$ term along with additional modifications to energy density if it contains inverse scale factors. Since we consider only a cosmological constant, the inverse scale factor modifications appear only via $A(v)$ term which is 
\begin{align}
    A(v) = - \frac{27 K}{4} \sqrt{\frac{8\pi}{6}} \frac{l_{Pl}}{\gamma^{\frac{3}{2}}}  |v| \left|\left|v-1\right|- \left|v+1\right|\right|,
\end{align}
with $K = 2/3\sqrt{3\sqrt{3}}$ and $v$ is related to physical volume $V$ as 
%
\begin{align}\label{Vphysical}
V =  \left(\frac{8\pi \gamma l_{Pl}^2}{6}\right)^{\frac{3}{2}} \frac{v}{K},
\end{align}
%
%

%

To derive the modified Friedmann equation, one needs to find the Hamilton's equation for $v$, which is given by
\begin{align}
\dot v = \{v, \mathcal{H}_{eff}\} = - \frac{\gamma \bar\mu A(v)}{2} \left(\frac{8\pi \gamma l_{Pl}^2}{6}\right)^{-1} K^{2/3} |v|^{1/3} \sin \bar\mu \left(c- k\right) \cos \bar\mu \left(c- k \right) .
\end{align}
To obtain physical solutions we need to demand that the effective Hamiltonian constraint vanishes. Using 
 $\mathcal{H}_{\mathrm{eff}} \approx 0$, one can eliminate dependence on variable $c$ and obtain the modified Friedmann equation containing both holonomy and inverse scale factor corrections as:
\begin{align}\label{effective}
 \nonumber H^2 = \left(\frac{\dot v}{3 v}\right)^2 & = \left(\frac{8\pi G}{3} {\rho} + {\tilde{A}(v)} \frac{k \chi}{\gamma^2 \Delta}\right)\left({\tilde{A}(v)} - \frac{{\rho}}{\rho_{c}} - {\tilde{A}(v)} k \chi\right)
 \\ &  =\frac{8 \pi G}{3} \left[{\rho} - \tilde{A}(v)\rho_{1}\right]\left[\frac{\tilde{A}(v)\rho_{2} - {\rho}}{\rho_{c}}\right],
\end{align}
where $\rho_{c}  = {3}/({8 \pi G \gamma^2 \Delta })$ is critical energy density, and 
\begin{equation}
\rho_{1} =  - \rho_{c} \, k \, \chi, ~~~
\rho_{2}  = \rho_{c} \left(1- k \chi\right),
\end{equation}
and absorbing some prefactors in $A(v)$, we define $\tilde{A}(v)$ as follows
\begin{align}\label{Atilde}
    \tilde{A}(v) = \frac{1}{2} \left|\left|v-1\right|- \left|v+1\right|\right|.
\end{align}
We see from the modified Friedmann equation that there are two turnarounds of the scale factor, the first at $\rho = \tilde A(v) \rho_1$ and the second at $\rho = \tilde A(v) \rho_2$. The nature of turnaround, whether it is a bounce or a recollapse can be determined using Raychaudhuri equation. When the initial conditions are such that universe evolves to a large macroscopic universe  the first turnaround corresponds to the classical recollapse of the universe while the second to the quantum bounce, but these can reverse if one considers a highly quantum universe \cite{Dupuy:2016upu}. In such a case, the quantum bounce occurs at the first turnaround and a quantum recollapse occurs at the second turnaround.

Moreover, one can also find the Raychaudhuri equation including both the holonomy and inverse scale factor corrections as follows (see Appendix \ref{Appendix A} for the derivation)
\begin{align}\label{Raychaudhuri-t}
\nonumber \frac{\ddot a}{a}& = - \frac{4 \pi G}{3}\left(  \left({\tilde{A}(v)} - 3 v \tilde{A}^{\prime}(v)  \right) \rho + 3 \tilde{A}(v) P \right) + \frac{16\pi G}{3} \left(\left(-\frac{1}{2} + \frac{3}{2} \tilde{A}(v)\right) \rho + \frac{3}{2}P\right)\left(\frac{\rho}{\rho_{c}} + {\tilde{A}(v)} k \chi\right)   \\&  \nonumber +  {\tilde{A}(v)} \frac{k\chi}{\gamma^2 \Delta}\left({\tilde{A}(v)} + 3 v \tilde{A}^{\prime}(v)   - \left(\tilde{A}(v)\right)^2\right)  \\ & \nonumber + \left(-1 + 2 {\tilde{A}(v)} - 3 \frac{v}{\tilde{A}(v)} \tilde{A}^{\prime}(v)\right) {\tilde{A}(v)} \frac{k\chi}{\gamma^2 \Delta} \left(\frac{\rho}{\rho_{c}} + {\tilde{A}(v)} k \chi \right) \\ &   +  \left[2  \left( {\tilde{A}(v)} - \left({\tilde{A}(v)}\right)^2\right) \frac{k \chi}{\gamma^2 \Delta}-2 {\tilde{A}(v)} \frac{k \zeta}{\Delta \gamma^2}\right] \left( \frac{\rho }{\rho_{c}} + {\tilde{A}(v)}k \chi - \frac{1}{2} {\tilde{A}(v)}\right),
\end{align}
where
\begin{align}
\zeta = \sin^2 \bar \mu -  \bar \mu \sin\bar \mu \cos \bar\mu.
\end{align}

\begin{figure}
    \centering
    \includegraphics[scale = 0.8]{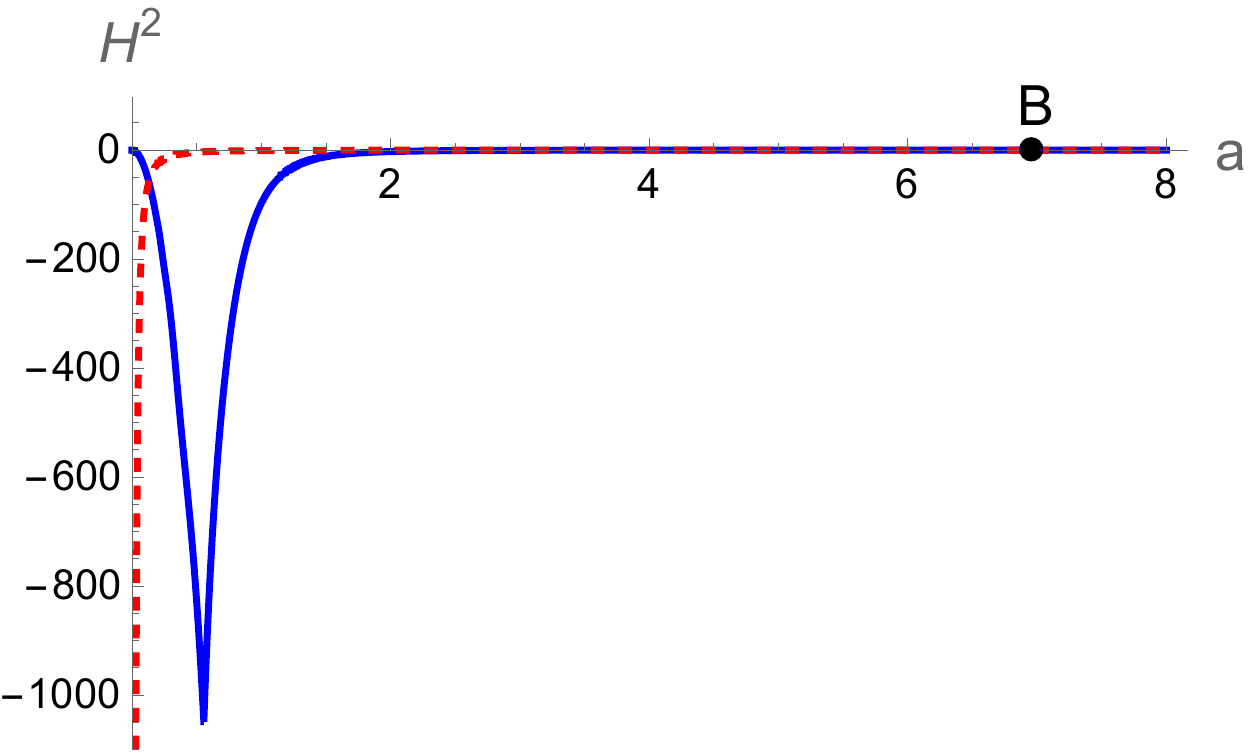}
    \caption{Square of Hubble parameter versus scale factor for $\Lambda = 0.1$. The red dashed line is for classical cosmology and the blue curve is for LQC with both holonomy and inverse scale factor corrections. Point B is the bouncing turnaround point for a large contracting universe as in the classical theory. Note that in classical theory $H^2$ and as a result the spacetime curvature diverges, whereas it vanishes in LQC.}
    \label{Hubble}
\end{figure}

One can easily check from Eq. (\ref{Atilde}) that for $v>1$, $\tilde{A}(v) \rightarrow 1$ and ${\tilde{A}}^{\prime}(v) \rightarrow 0$, and Eq. (\ref{Raychaudhuri-t}) reduces to the Eq. (32) in Ref. \cite{Singh:2010qa} if one considers just the holonomy correction. Given the modified Raychaudhuri equation, one can determine the nature of the turnaround points. In fact, the turnaround point is bouncing turnaround point if $\ddot a>0$, it is recollapsing turnaround point if $\ddot a < 0$ and it is an Einstein static  universe when $\ddot a=0$. 
In Fig. \ref{Hubble} we plot the square of the Hubble parameter versus scale factor in which the  dashed (red) line represents classical cosmology and the solid (blue) curve represents LQC with both holonomy and inverse scale factor corrections given in Eqs. (\ref{Friedmann}) and (\ref{effective}). One can see that the Hubble parameter diverges to infinity at zero scale factor which means there is a singularity at zero scale factor in classical cosmology. However, in case of LQC including both holonomy and inverse scale factor corrections, the Hubble parameter is zero at zero scale factor which means that $\dot a$ vanishes. One should note that the square of Hubble parameter is negative between zero and the classical  turnaround point `$B$' (which is a classical bounce for large contracting universe) which means it is classically forbidden region. Furthermore, from Eq. (\ref{Raychaudhuri-t}), one can see that in the small volume limit, i.e., $v\ll 1$, $\tilde{A}(v) \simeq v$, and as a result all the terms will be zero at right hand side of of Eq. (\ref{Raychaudhuri-t}) except the second term. But the second term is a constant for cosmological constant which implies that at vanishing scale factor $\ddot a$ vanishes.  Hence, one can conclude that at vanishing scale factor,  $\dot a= \ddot a=0$ in LQC in presence of both holonomy and inverse scale factor corrections. The universe is in a Einstein static like state albeit that it is a solution not of GR but of LQC. Thus, in this case the universe can start from zero size and zero energy, almost nothing, while the singularity is resolved due to quantum gravity effects. This will has important implication for tunneling wavefunction proposal as we discuss in next section by capturing quantum gravity effects in the effective minisuperspace potential.

\section{ Effective minisuperspace potential}\label{Section IV}

As we discussed in the introduction, tunneling and no-boundary proposals explain the boundary conditions for the universe, however, they do not resolve the singularity in a dynamical way. In fact, since the Lorentzian geometry is singular, Euclidean continuation is used to close off the geometry from bottom making it non-singular. In this section we are going to discuss what happens if one takes into account the quantum gravity effects which generally result in non-singular Lorentzian 
geometry. In doing so, having effective Friedmann equation for $k=1$ LQC, we read the effective minisuperspace potential capturing quantum gravity effects as it was discussed in Sec. \ref{Section II}. Then we will discuss the results for three cases: first, tunneling to classical expanding universe including just holonomy corrections, second, tunneling to classical expanding universe including both holonomy and inverse scale factor corrections, and finally tunneling to quantum cyclic universe when the cosmological constant is chosen to be large in Planck units. In the first two cases the value of the cosmological constant is such that the maximum of the energy density is not reached in LQC, but in the third case we choose a value of $\Lambda$ such that the latter is reached and there are two turnaround points -- a quantum  bounce and a quantum recollapse. Unlike the first two cases, the third case corresponds to a fully quantum universe both at the bounce and the recollapse points.

\begin{figure}[tbh!]
    \centering
    \includegraphics[scale = 0.7]{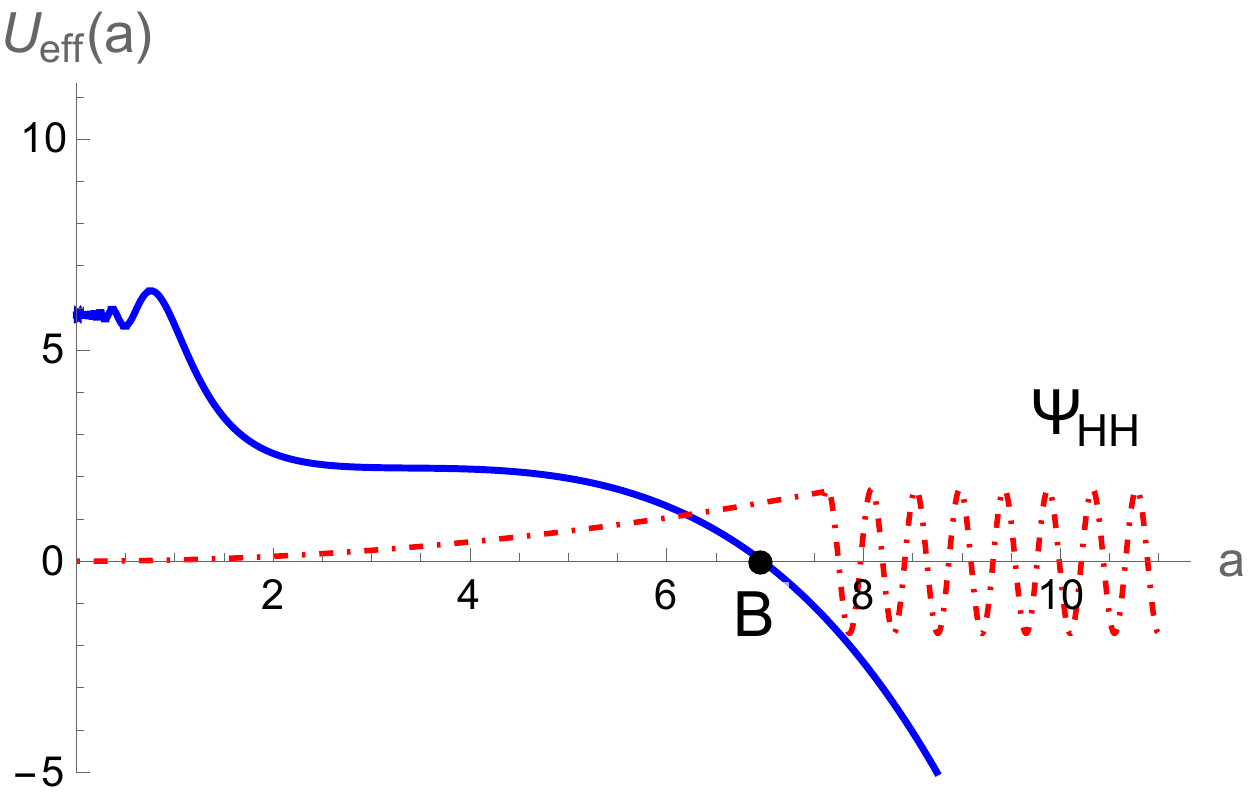}
    \caption{Effective minisuperspace potential and schematic behavior of wavefunction of the universe considering only holonomy corrections in LQC for $\Lambda = 0.1$ in Planck units. Point $B$ denotes the bouncing turnaround point for a large contracting classical universe.  }
    \label{fig2}
\end{figure}

\subsection{Non-tunneling to classical expanding universe without inverse scale factor corrections}

Let us consider the case of loop quantum cosmological model with only holonomy modifications. In this case i.e., $\tilde{A}(v)=1$, and one can read the effective minisuperspace potential from Eq. (\ref{effective}) as follows
\begin{align}\label{A(v)=1}
    U_{eff}(a) &= -\frac{12}{8 \pi G} a^4  \left( {{\rho}}- \rho_{1}\right)\Big(\frac{\rho_{2}-{{\rho}}}{\rho_{c}}\Big),
\end{align}
in which $\rho = \Lambda/(8 \pi G)$. In this case, the effective dynamics results in a quantum bounce when the classical singularity is approached. The effective minisuperspace potential in Eq. (\ref{A(v)=1}) reduces to effective minisuperspace potential obtained for Wheeler-DeWitt quantum cosmology in Eq. (\ref{V}) in the limit when the area gap $\Delta \rightarrow 0$.
 In Fig. \ref{fig2}, we illustrate the  effective minisuperspace potential in case of a small cosmological constant. One can see that the potential has classically allowed region only for scale factor larger than the classical bouncing turnaround point, i.e., $a> B$. Note that this point does not occur in the Planck regime but rather corresponds to the bounce of a large classical de Sitter universe as in the classical cosmology and Wheeler-DeWitt theory.  It should not be confused with the point of quantum bounce for general matter in LQC. 
 For this particular value of cosmological constant, the bouncing turnaround point, $B$, has different value from what obtained in the Wheeler-DeWitt quantum cosmology and that is due to quadratic term of energy density appearing in the effective Friedmann Eq. (\ref{effective}). The LQC and the Wheeler-DeWitt values for the classical bounce approach each other if one chooses a much smaller  cosmological constant. In such a case, the effective minisuperspace potential in Eq. (\ref{A(v)=1}) matches the  effective minisuperspace potential in Eq. (\ref{V}) at large scale factor. Note that in contrast to Fig. 1, the shape of potential (blue curve) has changed from barrier to step-like barrier near to zero scale factor due to quantum geometry effects. In other words, zero scale factor is forbidden in LQC for this case because of the quantum bounce and is not classically allowed unlike the case in Wheeler-DeWitt quantum cosmology. Furthermore, the height of the potential is also larger than the  effective minisuperspace potential in Eq. (\ref{V}) and which affects  the nucleation probability rate in Eq. (\ref{VHH}). From the behavior of eigenfunctions in loop quantization of $k=1$ model where holonomy modifications result in a quantum bounce one finds an exponential suppression of eigenfunctions for values of scale factor less than the bounce \cite{Ashtekar:2006es}.\footnote{This can also be shown analytically for an exactly solvable model in LQC \cite{craig}.}  Thus, the wavefunction in LQC must decrease towards the big bang under barrier (red dashed curve in above figure) in case of step-like potential. Hence, in this case the loop quantum universe cannot nucleate from nothing to classical expanding universe if the wavefunction is defined using the tunneling boundary conditions. In fact, in this case, the tunneling boundary conditions are not satisfied since the decaying mode cannot be dominant wave mode. So in this case, tunneling wave function proposal is incompatible with quantum gravity effects coming from just holonomy corrections. Rather, the allowed wavefunction would have features as in the no-boundary proposal. For this reason we have labeled this wavefunction as $``\Psi_{HH}"$ even though the above picture is different from the original Hartle and Hawking's no-boundary proposal.

 It is to be noted that above conclusion is valid for LQC in the situation when the holonomy effects resolve the singularity by a bounce. In the Wheeler-DeWitt case a scenario on similar lines which converts the barrier potential to a step like potential due to Casimir effect was studied in \cite{Mithani:2014toa}. In this case it was possible to impose boundary conditions with WKB approximations in such a way that one can obtain an increasing wave mode towards the zero scale factor under the step potential thus making tunneling wavefunction proposal viable.  Such a strategy can not work for LQC where the singularity at $a=0$ is never reached in presence of holonomy effects and the wavefunction is suppressed as one probes smaller scale factors than the bounce.

\subsection{Tunneling to classical expanding universe with inverse scale factor modifications}

\begin{figure}[tbh!]
    \centering
    \includegraphics[scale = 0.6]{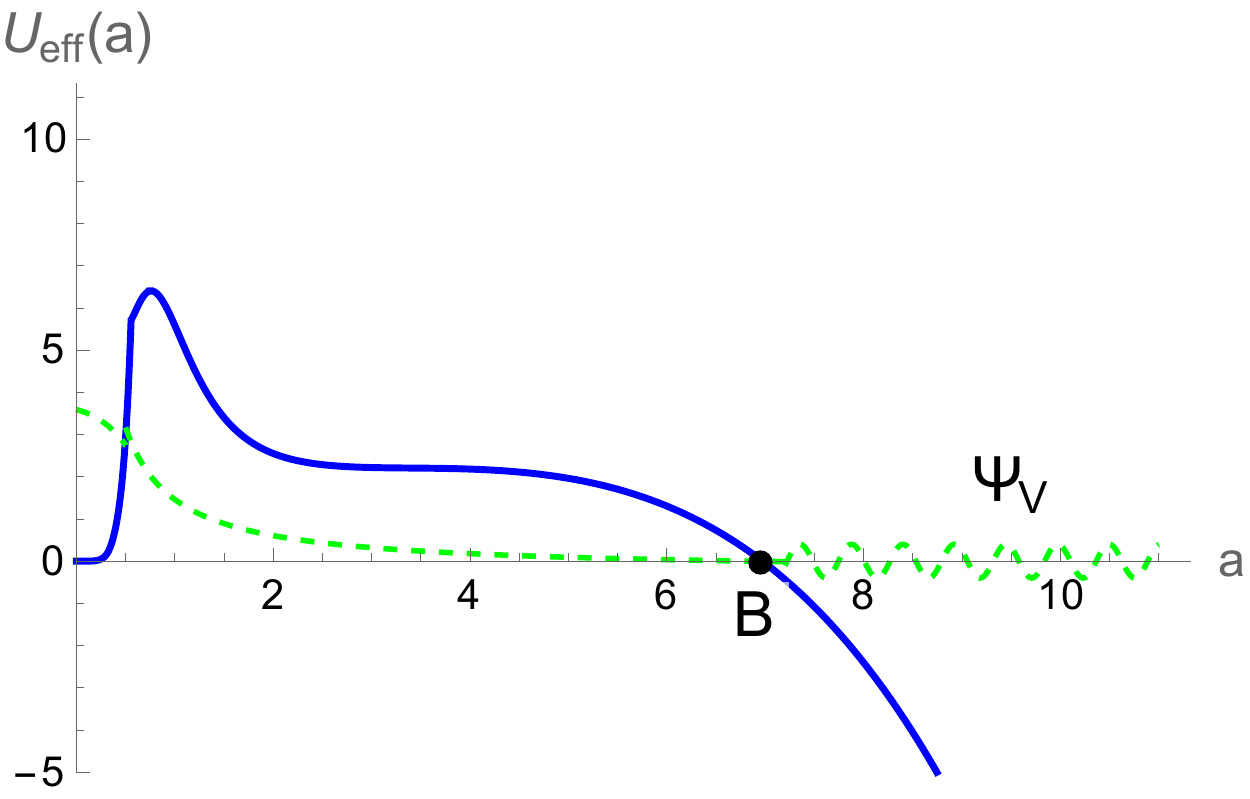}
    \includegraphics[scale = 0.6]{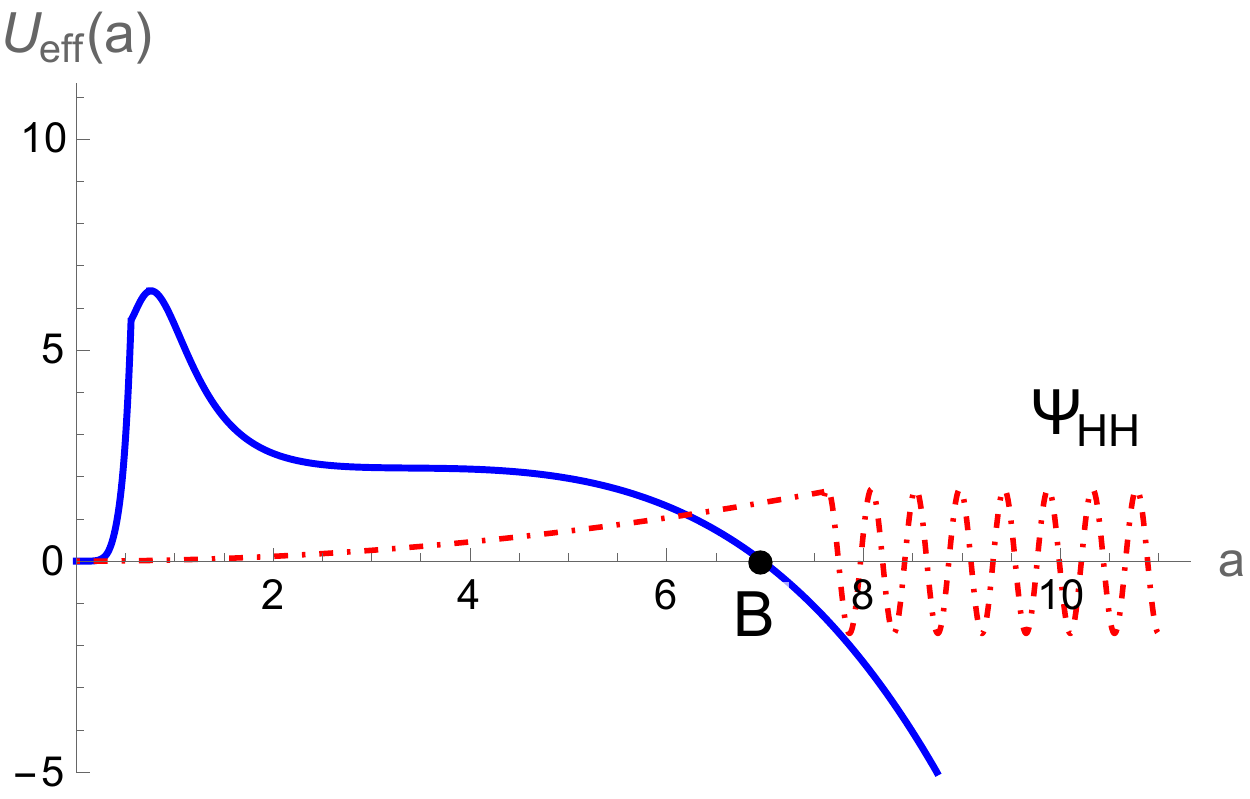}
    \caption{Schematic behavior of tunneling  (left) and no-boundary (right) wavefunction proposals considering both the holonomy and inverse scale factor corrections in LQC for $\Lambda =0.1$. Point $B$ denotes the classical bouncing turnaround point.}
    \label{fig3}
\end{figure}
Now we consider both the holonomy and inverse scale factor corrections in the effective potential which is now given by 
\begin{align}
    U_{eff}(a) &= -\frac{12}{8 \pi G} a^4  \left({{\rho}}- \tilde{A}(v)\rho_{1}\right)\Big(\frac{\tilde{A}(v)\rho_{2}-{{\rho}}}{\rho_{c}}\Big),
\end{align}
where it reduces to  effective minisuperspace potential Eq. (\ref{V}) in the limit $\tilde A(v) \rightarrow 1$ and $\Delta \rightarrow 0$.  In Fig. \ref{fig3}, we illustrate the effective minisuperspace for small cosmological constant, again choosing $\Lambda = 0.1$ in Planck units. In this case, one can see that there are two allowed regions for the   effective minisuperspace potential similar to Wheeler-DeWitt quantum cosmology. Due to singularity resolution at $a=0$ where $\dot a = \ddot a = 0$, there is an allowed region for single point at $a=0$. At large scale factors, since one recovers classical theory there is a classically allowed region also for $a > B$. We find that the effective minisuperspace potential (blue curve) recovers its barrier shape similar to Wheeler-DeWitt quantum cosmology taking into account both holonomy and inverse scale factor corrections. Note that the classical bouncing turnaround point has different value in comparison to the value in Wheeler-DeWitt quantum cosmology due to quantum geometry effect and also the height of the potential is larger than  effective minisuperspace potential in Eq. (\ref{V}). In this case, the universe can nucleate out of nothing to a classical expanding universe while the wavefunction of the universe is uniquely defined either by a tunneling wavefunction like scenario modified with quantum geometric effects (green curve in the left panel of Fig. \ref{fig3}) or a  no-boundary like scenario with LQC modifications (red curve in the right panel of Fig. \ref{fig3}). As we discussed in Sec. \ref{Section III}, the universe at zero scale factor is in a  Einstein static like universe with zero size while the singularity is resolved due to quantum geometry effects. Unlike the Wheeler-DeWitt theory, the spacetime curvature vanishes at this scale factor in LQC. Therefore, in this case one does not need to use a Euclidean continuation to resolve the singularity. In fact, quantum gravity effects coming from both holonomy and inverse scale factor corrections result in a non-singular version of tunneling wavefunction where singularity is resolved dynamically due to quantum gravity effects. However, one should point out that although quantum gravity effects may also result in non-singular version of no-boundary proposal, one needs to use Wick rotation in order to make the path integral convergent as it is the case in the original proposal.

\begin{figure}[tbh!]
    \centering
    \includegraphics[scale = 0.6]{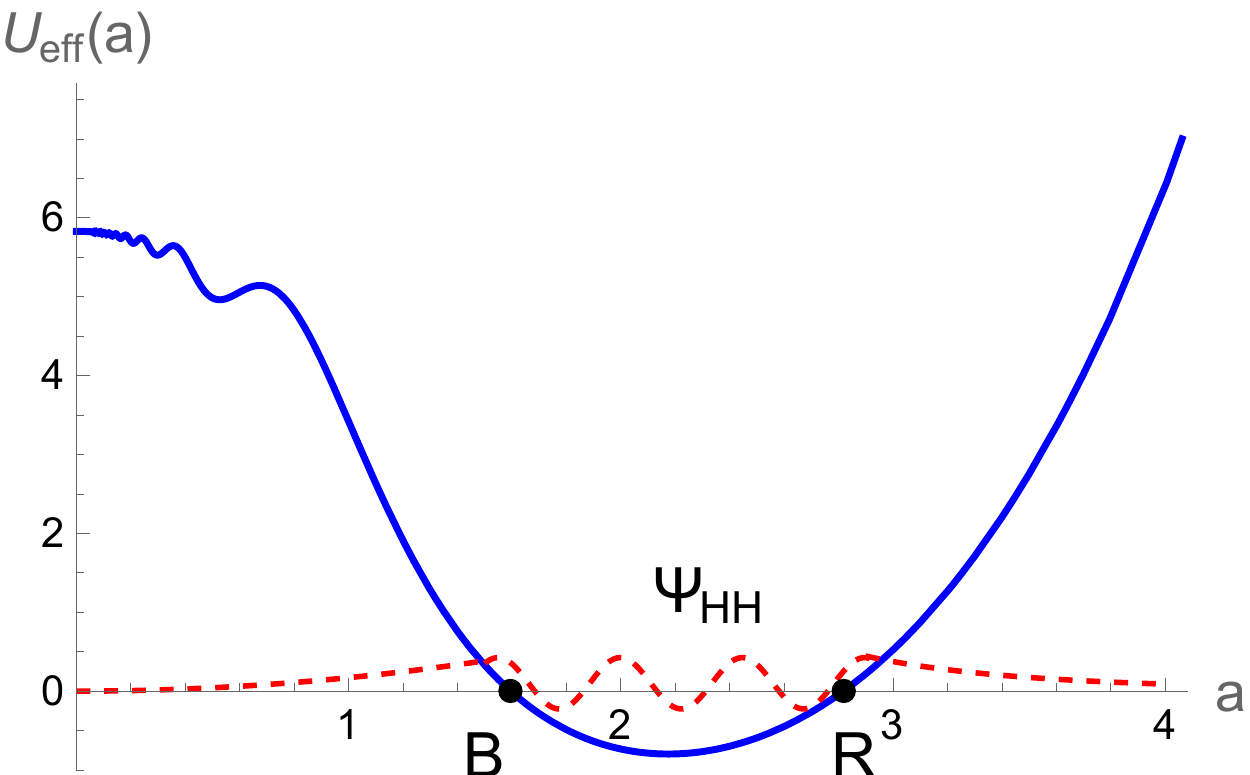}
    \caption{Schematic behavior of wavefunction of the universe considering just holonomy corrections in LQC for $\Lambda =12$. Points $B$ and $R$ denote the quantum bouncing and quantum recollapsing turnaround points where the energy density is in the Planck regime.}
    \label{fig4}
\end{figure}

\begin{figure}[tbh!]
    \centering
    \includegraphics[scale = 0.6]{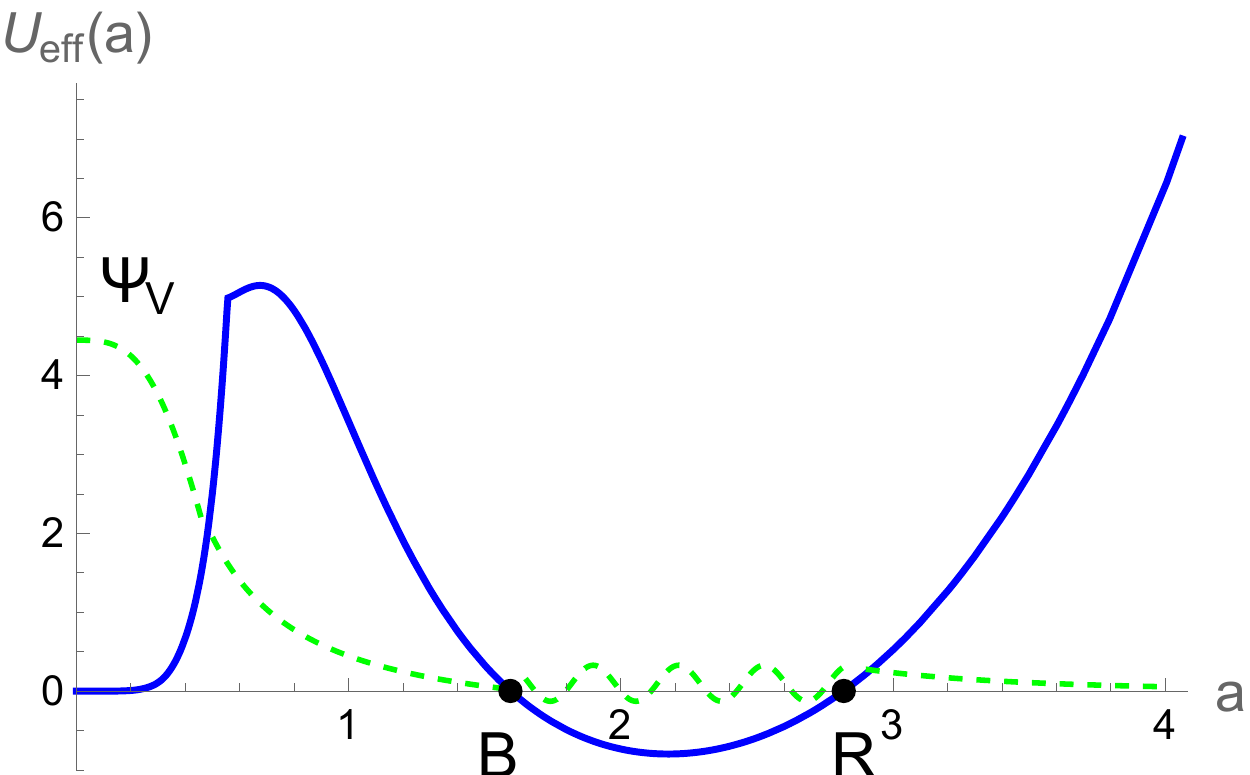}
    \includegraphics[scale = 0.6]{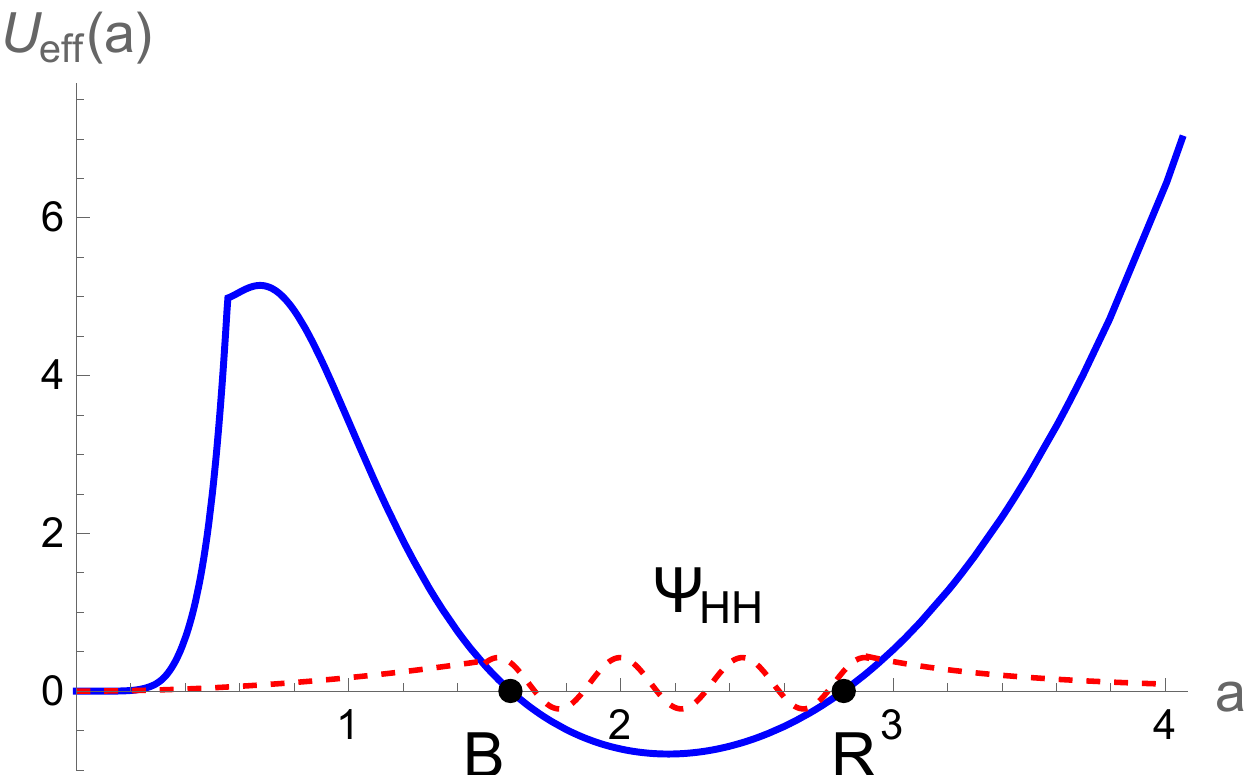}
    \caption{Schematic behavior of tunneling (left) and no-boundary (right) wavefunction proposals considering both the holonomy and inverse scale factor corrections in LQC for $\Lambda =12$. Points $B$ and $R$ denote the quantum bounce and quantum recollapse points.}
    \label{fig5}
\end{figure}

\subsection{Tunneling to a quantum cyclic universe}



Now we will consider the case that cosmological constant is large in Planck units.  Note that for a very large value of $\Lambda$, the second parenthesis in the modified Friedmann equation \eqref{effective} changes its sign. The allowed region in the classical regime becomes forbidden. In this case the universe will recollapse before it become classical, hence, the universe is always in a quantum regime. In Fig. \ref{fig4} we illustrate the effective minisuperspace potential for considering just holonomy corrections with a large cosmological constant. In this case, the universe has two distinct turnaround points, bouncing and recollapsing turnaround points. Hence, the universe can oscillate between these two turnaround points. Both of these turnaround points are quantum in nature. In particular, the  quantum bounce (denoted with `$B$') occurs at $ \rho = \rho_1$ and a quantum recollapse (denoted with `$R$') at $\rho = \rho_2$. For the considered value of cosmological constant, $\rho_1 \approx \rho_2 \approx 0.477$ in Planck units. Moreover, the shape of the  effective minisuperspace potential is again similar to step-like potential and it has similar height as in the case of a small cosmological constant. Therefore, there is only allowed region between the quantum bounce and quantum recollapse points and the universe can nucleate from nothing to quantum cyclic universe instead of a classical expanding universe. However, in this case, the wavefunction of the universe cannot be defined considering tunneling boundary conditions due to the step-like shape of potential near the zero scale factor. So we conclude that if one considers just the holonomy correction, tunneling wavefunction proposal is incompatible with quantum gravity effects. 

In  Fig. \ref{fig5}, we illustrate the effective minisuperspace potential in case of considering both holonomy and inverse scale factor corrections with large cosmological constant. Again, the effective minisuperspace has two distinct turnaround points and there is an allowed region between quantum bouncing and quantum recollapsing turnaround points. Since under the barrier either of decaying or growing mode can be chosen, one can conclude that the universe can tunnel from nothing to a quantum cyclic universe while the wavefunction of the universe is uniquely defined by either the tunneling (green curve in left panel of Fig. \ref{fig5}) or the  no-boundary (red curve in right panel of Fig. \ref{fig5}) conditions with quantum geometric modifications. The value of the cosmological constant at which a quantum cyclic universe can become a possibility is found to be governed by the area gap in quantum geometry. From the modified Friedmann dynamics, one finds that it is for $\Lambda > 3/\gamma^2 \Delta$ that one obtains such a cyclic behavior. Moreover, we should point out there is a possibility that universe also tunnel back to zero scale factor in this case as pointed out in the case of Wheeler-DeWitt quantization in Ref. \cite{Mithani:2011en}. In fact, as the universe recollapses and reaches the bounce turnaround point it can either bounce back or tunnel to zero scale factor, stay there and tunnel back to cyclic universe.\footnote{This result holds true in the presence of a massless scalar field if one takes into account inverse scale factor modifications to the energy density \cite{MP}.} 


\section{Summary} 

In GR,  the universe begins from the big bang singularity which is the boundary of the classical spacetime.  Several proposals have been put forward  to address the issue of boundary conditions of the universe among which the tunneling wavefunction proposal and the no-boundary wavefunction proposal are the most popular ones. These proposals attempt to define the boundary condition of the universe in such a way that the entire universe is self-contained explaining the initial conditions for the universe and the singularity is avoided using a Euclideanization procedure. However, these two proposals are based on the semi-classical aspects of gravity employing the same continuum differential geometry as the Wheeler-DeWitt theory and do not take into account non-perturbative quantum gravity effects. On the other hand, it is generally believed that the quantum spacetime resulting from non-perturbative quantum gravity will take us beyond the limitations of classical spacetime in GR as well as the Wheeler-DeWitt theory and provide new insights on the boundary conditions for the wavefunction of the universe.

In recent years, rigorous results from quantization of various cosmological spacetimes in LQC indicate that 
the big bang singularity is replaced by a big bounce \cite{Ashtekar:2011ni}. The underlying quantum dynamics in LQC can be understood using an effective spacetime description whose validity is assumed at all the scales in this work. This effective dynamics results in an effective Hamiltonian from which one can obtain a modified Friedmann dynamics. 
For spatially closed models in LQC, quantum geometry can result in two types of modifications for spatially closed models. One of these, which plays the dominant role in bounce,  arises because of holonomies of the connection taken over loops which have a minimum area determined by quantum geometry. Apart from the holonomy modifications in LQC there are also inverse scale factor modifications resulting from the underlying quantum geometry which become significant only at very small scale factors.  If one is interested in the backward evolution of a large macroscopic universe, quantum bounce happens  at a Planckian curvature scales but at scale factors much larger than the Planck length, then the inverse scale factor modifications remain subdominant. However, if one is interested in  obtaining the detailed picture of quantum spacetime, especially near the vanishing scale factor,  we need to include both the holonomy and inverse scale factor modifications. Since in a bouncing universe the vanishing scale factor is excluded one may be tempted to conclude that LQC is incompatible with the tunneling wavefunction proposal. The goal of this manuscript was to explore this issue in detail including both the holonomy and inverse scale factor modifications to understand the viability of tunneling wavefunction proposal.\footnote{There have been works investigating the no-boundary proposal in LQC but the viability of tunneling proposal was so far not  explored \cite{Brahma:2018elv, Bojowald:2018gdt, Brahma:2018kkr}.}

To investigate this issue we used the correspondence between Wheeler-DeWitt equation and Friedmann equation in one-dimensional minisuperspace. We used the fact that the  effective minisuperspace potential can be extracted from the modified Friedmann equation in LQC by an overall scaling by a term proportional to a power of scale factor. Then, deriving the   effective minisuperspace potential capturing quantum geometry effects, we found that the latter  modify the qualitative picture of creation of the universe out of nothing or tunneling wavefunction proposal in several different ways.  First, we find that if one considers just the holonomy corrections in LQC, the shape of the barrier changes to a step-like potential in the zero scale factor regime. Hence, zero scale factor regime is not allowed and the universe cannot tunnel from nothing to an expanding universe satisfying tunneling boundary conditions. However, when one takes into accounts holonomy and inverse scale factor corrections together, as it should be the case for spatially closed models in LQC, the  effective minisuperspace potential recovers its barrier shape as in Wheeler-DeWitt quantum cosmology. In fact we found that the zero scale factor is allowed where the universe is in Einstein static like phase. Therefore, we find  that the universe can tunnel from an Einstein static like phase with zero size to a classical expanding universe while the wavefunction can be uniquely defined using tunneling boundary conditions. In this case, zero scale factor is accessible while the singularity is resolved without complexifying the metric as it is the case in Vilenkin's original proposal. More precisely, quantum geometry effects with both holonomy and inverse scale factor corrections result in a non-singular version of tunneling wavefunction proposal without introducing any additional method to overcome the singularity. Second, we find that if the cosmological constant is small, the quantum gravity effects are negligible and the universe starts from infinite size, collapses, bounces back in the classical regime and re-expands similar to as in classical cosmology. However, when the cosmological constant is large the energy density of the universe can reach the maximum energy density for which the universe recollapses, as result of which the universe will go through a cyclic evolution.  In the latter case, the universe experience two turnarounds -- one quantum bounce and another as a quantum recollapse.  The critical value of cosmological constant above which this behavior occurs is determined by the the quantum geometry. Hence, the universe can tunnel from nothing at the vanishing scale factor to either a classical expanding universe  or a quantum cyclic universe in pure de Sitter cosmology in LQC depending on the value of cosmological constant. Finally, we also found that the height of the barrier is larger when one takes into account quantum gravity effects which can change the  rate for nucleation probability. However, what is important is that the universe will tunnel through the barrier no matter how small this probability rate is. An open question in this analysis is the way these predictions get affected if one considers a full quantum treatment using the quantum Hamiltonian constraint in LQC rather than the effective dynamics. Such an analysis is expected to yield further insights on the viability of tunneling wavefunction boundary conditions from the perspective of the physical Hilbert space in LQC.

\begin{acknowledgements}
This work is supported by the NSF grant PHY-2110207.
\end{acknowledgements}

\appendix 
\section{Derivation of Raychaudhuri equation including $A(v)$ term}\label{Appendix A}

In this appendix, we derive the Raychaudhuri equation including both holonomy and inverse scale factor terms. Starting with Hamiltonian constraint Eq. (\ref{constraint}), one can find the Hamilton's equation for $p$:
\begin{align}\label{dotp}
\dot p = \{p, \mathcal{H}_{eff}\} =-  \frac{\gamma \bar \mu}{3} A(v)\sin (\bar \mu (c-k)) \cos (\bar \mu(c-k)).
\end{align}
Taking time derivative from Eq. (\ref{dotp}), we have
\begin{align}
\nonumber \ddot p &=\left( \frac{\gamma \bar \mu}{6} \frac{\dot p}{p} A(v) - \frac{\gamma \bar \mu}{3} A^{\prime}(v) \dot v\right)\sin (\bar \mu (c-k)) \cos (\bar \mu(c-k)) \\& - \frac{\gamma \bar \mu}{3} A(v)\left(\dot{\bar \mu} (c-k) + \bar \mu \dot c\right) \left[\cos^2 (\bar \mu(c-k)) - \sin^2 (\bar \mu(c-k)) \right] \label{p1} \\ & = \left(2 +{3} \frac{v }{\tilde{A}(v)} \tilde{A}^{\prime}(v)- {\tilde{A}(v)}+ 2 {\tilde{A}(v)} \sin^2 \bar \mu (c-k)\right) 2 p H^2 + \frac{2 \sqrt{p}}{\gamma}{\tilde{A}(v)}\dot c \left[1- 2\sin^2 (\bar \mu(c-k)) \right], \label{p2}
\end{align}
where prime means derivative with respect to $v$ and we used $A(v) = - \frac{27 K}{2} \sqrt{\frac{8\pi}{6}} \frac{l_{Pl}}{\gamma^{\frac{3}{2}}}  |v| \tilde{A}(v)$ to reach the last equation. Using Eq. (\ref{Vphysical}), one can find that 
\begin{align}\label{ddota}
\nonumber \frac{\ddot a}{a} &= \frac{\ddot p}{2p} - H^2 \\ &= \left(1 + {3} \frac{v }{\tilde{A}(v)} \tilde{A}^{\prime}(v)- {\tilde{A}(v)}+ 2 {\tilde{A}(v)}\sin^2 \bar \mu (c-k)\right) H^2 + {\tilde{A}(v)} \frac{\dot c}{\gamma \sqrt{p}} \left[1- 2\sin^2 (\bar \mu(c-k)) \right].
\end{align}
Therefore, we need to find $\dot c$ to determine the modified Raychaudhuri equation. Using the Hamilton's equation for $c$, we get 
\begin{align}
\dot c &= \{c, \mathcal{H}_{eff}\} = \frac{8 \pi G\gamma}{3} \frac{\partial \mathcal{H}_{eff}}{\partial p}  \\& \nonumber =\frac{\gamma}{6} A^{\prime}(v) \left[\sin^2(\bar \mu (c-k)) - k \chi \right] \frac{\partial v}{\partial p}  + 4 \pi G \gamma \sqrt{p} \frac{\partial  \mathcal{H}_{M}}{\partial V}  \\ & + \frac{\gamma}{6} A(v) \left( 2 (c-k){\sin \bar \mu (c-k) \cos\bar \mu (c-k)}  - 2 k \sin\bar \mu \cos \bar\mu + 2 k \bar \mu (1+\gamma^2) \right) \frac{\partial \bar\mu}{\partial p}  \\ & = \gamma \sqrt{p} \left[-4\pi G  {\rho} \left(1 + \frac{v }{\tilde{A}(v)} \tilde{A}^{\prime}(v) \right)- 4 \pi G  P   +  H^2 + {\tilde{A}(v)} \frac{1}{\gamma^2 \Delta} \left(-k\chi + k\zeta \right)\right] \label{dotc}.
\end{align}
Moreover, using Hamiltonian constraint $\mathcal{H}_{eff} \approx 0$, we obtain
\begin{align}\label{sin}
\sin^2 \bar\mu (c-k) = \frac{1}{\tilde{A}(v)} \frac{{\rho}}{\rho_{c}} + k \chi
.
\end{align}
Then, using Eqs. (\ref{ddota}), (\ref{dotc}) and (\ref{sin}), one finds 
\begin{align}\label{ddota2}
\nonumber\frac{\ddot a}{a} & =  \left(1 +{3} \frac{v }{\tilde{A}(v)} \tilde{A}^{\prime}(v)- {\tilde{A}(v)} + 2  \left({ }\frac{\rho}{\rho_{c}} + \tilde{A}(v)k \chi \right)\right)  \left(\frac{8\pi G}{3} \rho + {\tilde{A}(v)}\frac{k \chi}{\gamma^2 \Delta}\right) \\ & \nonumber \times\left({\tilde{A}(v)}- \frac{\rho}{\rho_{c}} - {\tilde{A}(v)} k \chi\right) +  \left({\tilde{A}(v)}-  2\frac{\rho }{\rho_{c}}- 2 {\tilde{A}(v)}k \chi \right) \times \left[-4\pi G  {\rho} \left(1 + \frac{v }{\tilde{A}(v)} \tilde{A}^{\prime}(v) \right) \right. \\& \left. - 4 \pi G P   +  \left(\frac{8\pi G}{3} \rho + {\tilde{A}(v)}\frac{k \chi}{\gamma^2 \Delta}\right)\left({\tilde{A}(v)}- \frac{\rho}{\rho_{c}} - {\tilde{A}(v)} k \chi\right)  + {\tilde{A}(v)} \frac{1}{\gamma^2 \Delta} \left(-k\chi + k\zeta \right)\right].
\end{align}
After some simplification, one reaches the following modified Raychaudhuri equation including both holonomy and inverse scale factor corrections
\begin{align}\label{Raychaudhuri}
\nonumber \frac{\ddot a}{a}& =  - \frac{4 \pi G}{3}\left(  \left({\tilde{A}(v)} - 3 v \tilde{A}^{\prime}(v)  \right) \rho + 3 \tilde{A}(v) P \right) + \frac{16\pi G}{3} \left(\left(-\frac{1}{2} + \frac{3}{2} \tilde{A}(v)\right) \rho + \frac{3}{2}P\right)\left(\frac{\rho}{\rho_{c}} + {\tilde{A}(v)} k \chi\right)   \\&  \nonumber +  {\tilde{A}(v)} \frac{k\chi}{\gamma^2 \Delta}\left({\tilde{A}(v)} + 3 v \tilde{A}^{\prime}(v)   - \left(\tilde{A}(v)\right)^2\right)  \\ & \nonumber + \left(-1 + 2 {\tilde{A}(v)} - 3 \frac{v}{\tilde{A}(v)} \tilde{A}^{\prime}(v)\right) {\tilde{A}(v)} \frac{k\chi}{\gamma^2 \Delta} \left(\frac{\rho}{\rho_{c}} + {\tilde{A}(v)} k \chi \right) \\ &   +  \left[2  \left( {\tilde{A}(v)} - \left({\tilde{A}(v)}\right)^2\right) \frac{k \chi}{\gamma^2 \Delta}-2 {\tilde{A}(v)} \frac{k \zeta}{\Delta \gamma^2}\right] \left( \frac{\rho }{\rho_{c}} + {\tilde{A}(v)}k \chi - \frac{1}{2} {\tilde{A}(v)}\right) 
\end{align}
It is easily checked that this equation has the correct classical limit in the regime when volume is much greater than Planck volume and $\rho \ll \rho_c$.

 \end{document}